# UPGRADES OF BEAM DIAGNOSTICS IN SUPPORT OF EMITTANCE-EXCHANGE EXPERIMENTS AT THE FERMILAB A0 PHOTOINJECTOR

A.H. Lumpkin, A.S. Johnson, J. Ruan, J. Santucci, Y.-E Sun, R. Thurman-Keup, and H. Edwards
Fermilab, Batavia, IL U.S.A. 60510

*Abstract*

The possibility of using electron beam phase space manipulations to support a free-electron laser accelerator design optimization has motivated our research. An ongoing program demonstrating the exchange of transverse horizontal and longitudinal emittances at the Fermilab A0 photoinjector has benefited recently from the upgrade of several of the key diagnostics stations. Accurate measurements of these properties upstream and downstream of the exchanger beamline are needed. Improvements in the screen resolution term and reduced impact of the optical system's depth-of-focus by using YAG:Ce single crystals normal to the beam direction will be described. The requirement to measure small energy spreads (<10 keV) in the spectrometer and the exchange process which resulted in bunch lengths less than 500 fs led to other diagnostics performance adjustments and upgrades as well. A longitudinal to transverse exchange example is also reported.

--------------------------------

*Work supported by U.S. Department of Energy, Office of Science, Office of High Energy Physics, under Contract No. DE-AC02-06CH11357.

Keywords: optical transition radiation, YAG:Ce crystal and powder scintillators, emittance exchange
PACS: 41.60.Ap, 41.60.Cr

# I. INTRODUCTION

It is recognized that beam manipulations such as a flat beam transformation followed by an emittance exchange (EEX) could support a high gain free-electron laser (FEL) push to shorter wavelengths [1,2]. An ongoing program demonstrating the exchange of transverse horizontal and longitudinal emittances at the Fermilab A0 photoinjector (A0PI) addresses the latter of these beam manipulations. Recent upgrades to key optical diagnostics stations have improved our capabilities to measure EEX. The experiments rely on accurate measurements of the emittance properties upstream and downstream of the exchanger beamline. At gamma ~30 the nominal transverse beam sizes of 1 mm were not an imaging challenge. However, resolution limits are approached with the use of an array of 50-micron wide slits that sample the transverse phase spaces in order to measure divergences of less than 100 microradians. Such low divergences are evidenced by typically 20 times smaller images than the beam size. The 1-mm spacing of the slits also resulted in images with positions distributed over several mm which involved the depth-of-focus limits of the initial optics system. Improvements in the screen resolution term and reduction of the system depth-of-focus impact by using YAG:Ce single crystals normal to the beam direction will be described.

On the longitudinal side, the requirement to measure small energy spreads (<10 keV) in the spectrometer and bunch lengths less than 500 fs impacted the corresponding diagnostics performance specifications as well. Upgrades to the Hamamatsu C5680 streak camera and the addition of a Martin-Puplett interferometer addressed the short bunch lengths generated by the exchange process. An example of the EEX results obtained with the upgraded diagnostics will be presented.

# II. EXPERIMENTAL ASPECTS

The tests were performed at the Fermilab A0 photoinjector facility which includes an L-band photocathode (PC) rf gun and a 9-cell superconducting radiofrequency (SCRF) accelerating structure which combine to generate up to 16-MeV electron beams [3,4]. The Nd glass drive laser operates at 81.25 MHz with the micropulse structure counted down to 1 MHz. The frequency-quadrupled component at 263 nm is used to irradiate the $Cs_2Te$ PC. Due to the low electron-beam energies (14-16 MeV), one radiation converter strength, and/or slit sampling, we typically summed over micropulses in such cases to obtain adequate signal strength. The typical beam parameters for the initial conditions used in these experiments are listed in Table I. The 250-pC micropulse charge was chosen as a compromise between possible space-charge effects and diagnostic signal strength.

**TABLE I.** Initial beam parameters of the A0PI facility.

| Parameter | Value | Units |
|---|---|---|
| Energy | 14.3 ± 0.1 | MeV |
| Micropulse charge | 250 ± 20 | pC |
| Beam size ($\sigma$) | 1-2 | mm |
| Divergence ($\sigma$) | 50-100 | μrad |
| Bunch length ($\sigma$) | 2.9 ± 0.3 | ps |
| Emittance (norm.) | 2.6 ± 0.3 | mm mrad |

Fundamental to the observations of EEX were the careful beam transport and characterization of the transverse and longitudinal emittances before and after the process. The A0PI beamlines provided these capabilities through a set of rf beam position monitors (BPMs) and optical diagnostics stations as shown in Fig. 1. These imaging stations (crosses denoted with an "X#") use optical transition radiation (OTR) and/or scintillator converter screens. Since the initial report of observation of EEX [5], we have upgraded all of the divergence and spectrometer stations to dramatically reduce the magnitude of needed corrections to the simple image sizes that relate to beam size, divergence, or energy spread depending on the diagnostics configuration. All data were obtained with a micropulse charge of 250 pC, but in some cases 10-50 micropulses were used in a single macropulse integrated by the camera CCD chip. In all cases the photocathode dark current was subtracted by shuttering the drive laser to obtain a background image file. The former YAG:Ce 50-μm thick powder screens on an Al substrate oriented at 45 degrees to the beam direction [6] (with updated spatial resolution term in Section III) have been now replaced by 100-μm thick YAG:Ce single crystals oriented normal to the beam direction followed by a 45 degree mirror to direct the radiation to the optical system. This configuration would reduce the screen resolution term to less than 10 μm rms based on evaluation of previous reports [7,8] and also eliminate the depth-of-focus issue of the 45 degree scintillator for multiple slit images spread across several mm of the field of view. Energizing a downstream horizontal bending dipole sends the beam into a final beam dump and provides a spectrometer capability when combined with imaging of the beam size at the XS3 dispersive point in the focal plane. (The depth-of-focus effect was subsequently also addressed for beam image location in the direction perpendicular to the energy-dispersive direction in the spectrometers by the same configuration change.) The dispersion in x is 324 mm, and the CCD calibration factor is 47.5 μm per pixel. This means a 0.1 % energy spread would correspond to an ~324-μm beam size after corrections are done.

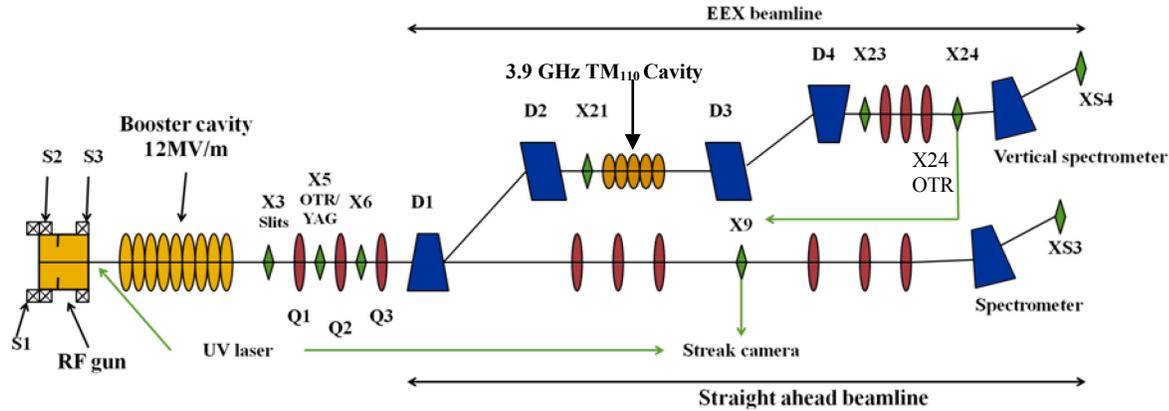

Figure 1: A schematic of the A0 photoinjector test area showing the PC rf gun, 9-cell booster cavity, transverse emittance stations, the OTR stations, the streak camera, and the EEX beamline with the two doglegs, 5-cell deflecting mode cavity, transverse emittance stations, and spectrometer.

The initial emittance sampling station was chosen at X3, and an optical transport system using a 50-mm diameter field lens and a flat mirror brought the light to the FireWire digital CCD camera with C-mount lens. In addition, a larger drift length for the initial divergence measurements was attained by using the X05 and X06 stations (drifts of 0.79 and 1.29 m, respectively) instead of X04 and X05 to view the slit images whose sizes were now dominated by the divergence-term contribution. The camera edge resolution function terms were evaluated with standard line-pair patterns and were typically about 40-45 μm, and the rms contribution of the finite slit width size was evaluated as 14 μm. These were subtracted in quadrature from the observed slit image profile sizes. So the projected beam sizes were recorded at X3 using OTR, and then vertical and horizontal slit assemblies were inserted in sequence at this position per the slits emittance measuring technique [9]. These slit assemblies were 3-mm thick tungsten plates separated with spacers to form 50-μm wide slits. The depth of metal was chosen to provide strong opacity to the 14.3-MeV beam. A MATLAB-based program calculated the emittances and Courant-Snyder (C-S) parameters (α,β,γ) based on the X3-X5 and X3-X6 image pairs [10]. Alternatively, the 4-dipoles (D1-D4) of the emittance exchange (EEX) line could be powered and emittance measurements done at an OTR Cross X23 after the fourth dipole with a drift of 0.56 m to Cross X24. The latter has both OTR and YAG:Ce crystal screen options. The OTR converter was an Al-coated optics mirror with a 1.0-mm thick glass substrate, and was mounted with its surface at 45 degrees to the beam direction on a stepper assembly. The X24 assembly provided vertical positioning with an option for a YAG:Ce crystal scintillator position. The charge was monitored by an upstream current monitor.

### III. SYSTEM RESOLUTION EFFECTS

There are several contributions to the observed beam image size. When they are uncorrelated, the terms can be added in quadrature as described by Lyons [11]. The terms that have been considered are: actual beam size (Act), camera resolution (cam), YAG:Ce powder screen effects (YAG), and the finite slit width (slit) as shown here:

$$Obs^2 = Act^2 + YAG^2 + cam^2 + slit^2 \quad , \tag{1}$$

and after solving for the actual beam size we have,

$$\text{Act} = [\text{Obs}^2 - \text{YAG}^2 - \text{cam}^2 - \text{slit}^2]^{1/2} . \tag{2}$$

In the spectrometer, the slit width is replaced by the beam size without dispersion, $\beta_x \varepsilon_x/\gamma$ (where $\gamma$ is the Lorentz factor and $\varepsilon_x$ is the normalized x emittance), and the actual beam size is now the dispersive term-energy spread product ($\eta_x \sigma_E$), and this gives:

$$\text{Obs}^2 = \text{Act}^2 + \text{YAG}^2 + \text{cam}^2 + \beta_x \varepsilon_x/\gamma , \tag{3}$$

and after solving for the actual beam size we have,

$$\text{Act} = [\text{Obs}^2 - \text{YAG}^2 - \text{cam}^2 - \beta_x \varepsilon_x/\gamma]^{1/2} = \eta_x \sigma_E . \tag{4}$$

### A. Scintillator screen resolution term

One of the main characteristics of powder scintillator screens in the past has been a limiting resolution term that was screen-thickness dependent such as found in $Al_2O_3$:Cr samples [12,13]. Another aspect involves the grain size of the scintillator particles since this should also contribute to the limiting resolution term. More recent versions of thin YAG:Ce powder screens utilized at our linac involved reported grain sizes at the 5-μm level [14]. However, we report our comparisons with OTR screen results that indicate that the 50-μm thick layer deposited on a 1-mm substrate and oriented at 45 degrees to the beam direction has a 1-sigma resolution term much larger than the grain-size value. Our investigations were limited by the nominal beam sizes available at this 16-MeV photoinjector so slit images and as well as focused beam images were used to provide a source term in one transverse dimension small enough to detect these effects. The studies were also complicated by the roughly 100 times stronger signal of the scintillator compared to the OTR screen for a given charge, and the effects of the RF amplitude and phase flatness over a pulse train. Our initial comparison of the powder screen and OTR in the same geometry is shown in Table II. For comparison purposes we also show the OTR polarization data in the table. First, it is noted that there is evidence for an OTR polarization effect where one uses the perpendicular polarization component to assess a given beam size dimension [15,16] in rows 1,2, and secondly no polarization effect observed as expected in the YAG:Ce based beam sizes at rows 3,4. Only one bunch was used with the brighter scintillator, so we would be reducing any macropulse effects in the beam in this case. On average, the vertically polarized OTR band images in rows 2 and 6 have a size of 101 μm compared to the total OTR image size average of 124.5 μm and the YAG:Ce image size average of 130 μm. These results are consistent with our previous hypothesis that the YAG:Ce powder screen has a limiting resolution term to be addressed. If we accept the OTR polarized data as the reference, then the implied YAG:Ce X5 screen term in quadrature would be 80±10 μm. The central value would be 40 μm if the total OTR image size were used as reference. In a second series of tests the powder screen images were compared to those of a YAG:Ce single crystal with surface normal to the beam direction and a resulting estimate of 50-60 μm for the powder term was obtained. We averaged these three results to obtain our resolution value of 60±20 μm for the present screens obtained from DESY [14]. As further context, we collect the results of several reports in the literature [12,13,17] where the powder screen beam image sizes were compared directly to either those of OTR or a YAG:Ce single crystal. Again we evaluated the differences in observed beam image sizes by postulating there was a powder screen term that could be treated in quadrature with the actual beam size as in Eq.1. As shown in Fig. 2, there is a strong indication of the thickness effect in these merged examples of data. Our present YAG:Ce result using the screen thickness at 45 degrees is in basic agreement with the deduced result of 40 μm at 60 MeV with 500 pC charge and for an assumed similarly deposited YAG:Tb screen, albeit used with its surface *normal* to the beam direction with light collected at the back angle via an annular metallic lens [17]. More comprehensive studies of YAG:Ce powder screens' yield, resolution, thickness effects, etc. have been done in the x-ray imaging community

[18], but the relativistic electron beams are more penetrating than the 100 keV x rays typically used so we can not directly apply the x-ray results in all parameters. As stated earlier, we have now changed to the single crystal screens with sub-10-µm resolution that basically do not have grain size or light scattering issues of the same type as that of the powder versions of YAG:Ce.

**TABLE II.** A comparison of beam image sizes using YAG:Ce powder and OTR screens, both oriented at 45 degrees to the beam at station X5. The YAG:Ce powder data used only 1 bunch as denoted by 1-Y in col. 4. Averages were based on fits to 10 images, and the variances of values of the fits were determined.

| X5 band | X5 Pol. | Laser Pol. | No. bunches | Fit sigma (pixels) | Size (µm) |
|---|---|---|---|---|---|
| V | No | P | 10 | 5.42±0.03 | 123.0 |
| V | V | P | 10 | 4.00±0.04 | 90.8 |
| V | No | P | 1-Y | 5.67±0.09 | 128.7 |
| V | V | P | 1-Y | 5.71±0.04 | 129.6 |
| V | No | S | 10 | 5.55±0.02 | 126.0 |
| V | V | S | 10 | 4.95±0.10 | 112.4 |

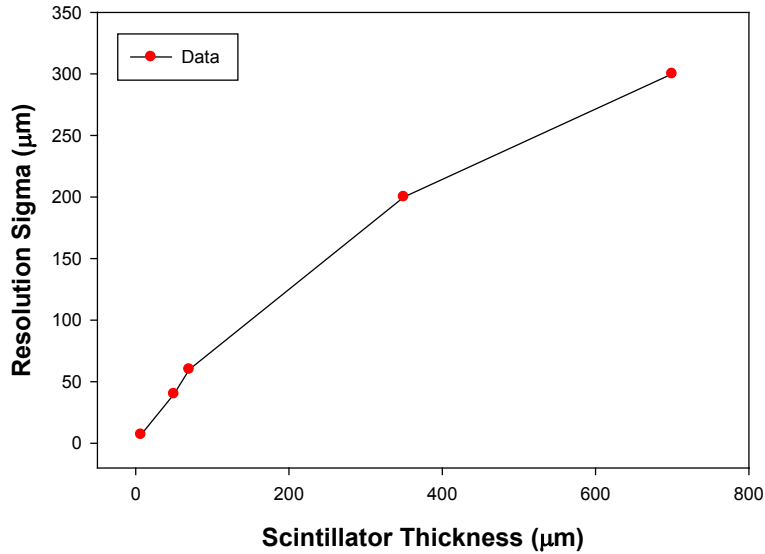

Figure 2: Combined plot of powder screen thicknesses and deduced spatial resolution terms based on a quadrature analysis of the reported observed image sizes with the OTR or reference image size. The lowest resolution value datum was estimated for a thin powder screen at the 5-µm grain size in thickness and at 45 degrees. The lines between points are used to guide the eye.

## B. Depth-of-focus effects

In the case of the divergence measurements based on the slit sampling, we have determined that the depth of focus of the optical arrangement was nonideal. If one samples the phase space with 1-mm slit spacing, the outer slit images are blurred by the degraded system resolution. This effect is graphically illustrated in Fig. 3, where we see that the observed slit-image size variation across the field of view can be reproduced by adding the z-dependent optical-bench-based camera-resolution term in quadrature with the minimum slit image seen at the focus for the X5 case. We see that from the central in-focus reference position for X5, even the next slit image position has a detectable growth. An even larger effect was seen at X24 where the slit-image spacings were 3 mm. To address such depth-of-focus effects in A0PI stations, we have installed at X5 and X24 a YAG:Ce single crystal with its surface plane normal to the beam direction and with a 45-degree mirror just downstream which directs the light to the optics. The resulting slit images were more uniform in size across the scene, and averaged divergences (and thus calculated emittances) were lower in subsequent tests.

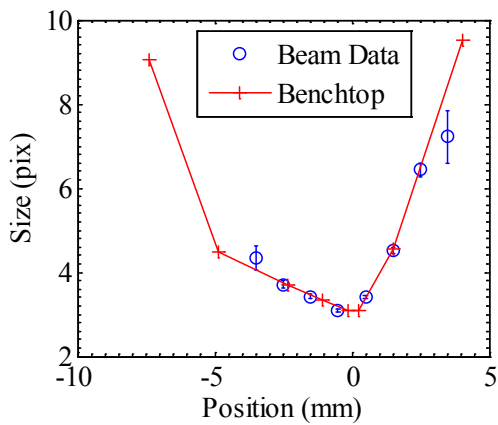

Figure 3: Comparison of the observed vertical slit-image sizes and calculated sizes from the bench tests depth-of-focus data added in quadrature with the minimum beam size at focus for X5.

## IV. EXPERIMENTAL RESULTS

Following our implementation of the diagnostic station upgrades, we returned to the basic EEX experiments. Example beam images are shown in Fig. 4a,b for the screens X3 and X5. The beam image and vertical slit image profiles were fit to Gaussian shapes as shown in Figs. 3 c,d. Typical full image sizes were 1-2 mm, but the slit images can be as small as 80 μm corresponding to ≈ 50 μrad divergence.

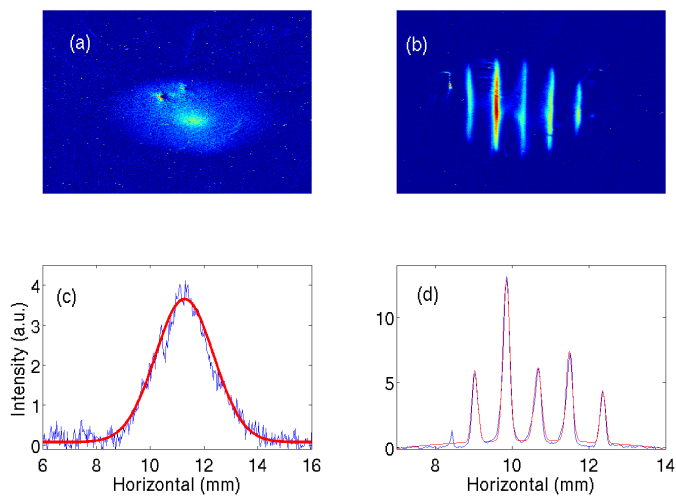

Figure 4: Composite display of a) X3 OTR image, b) X5 YAG:Ce crystal slit images, c) projected x profile (blue) and its Gaussian profile fit (red curve) and d) slit image x profiles with the Gaussian fitted curves (red).

The evaluation of phase space can be visualized by plotting the calculated divergences for each slit image sample and constructing the 2-D x-x' phase space [10] as shown in Fig. 5. In addition, from such data the Courant-Snyder parameters ($\alpha,\beta,\gamma$) were also determined and used as input for the emittance exchange model. Additional X3-X5 results are shown in Fig. 6. The partition of beam size (red circles) and divergence (blue triangles) varied with main solenoid (S2) current while the x emittances (black circles) were calculated as 2.0-2.6 mm mrad. The lowest divergence numbers result from large fractional corrections for the camera resolution term. The upgraded X6 station with a larger drift distance from the X3 station, and hence larger divergence-related effects, will ameliorate this issue as will be separately reported [19].

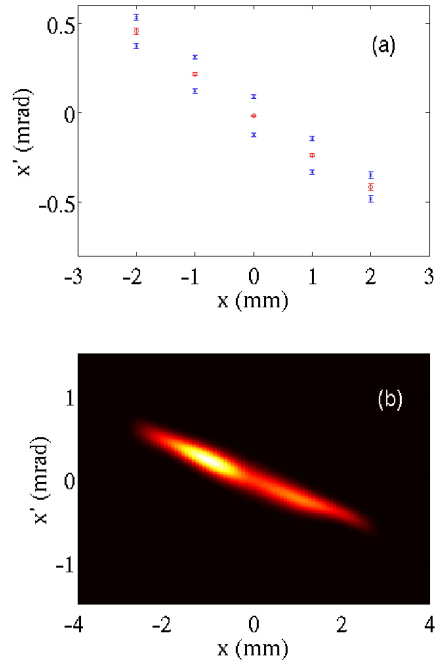

Figure 5: Plots of (a) the measured x-x' phase space and b) visualization of transverse phase space based on the slit sampling of the beam cross section at X3 and the slit images being observed at X5 as shown in Fig.4.

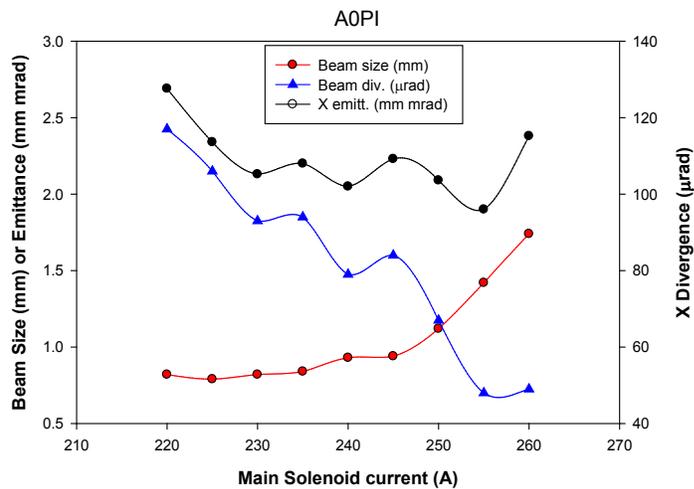

Figure 6: The variation of beam size (red circles), divergence (blue triangles), and normalized emittances (black circles) at X3-X5 versus the main solenoid current.

## V. EMITTANCE EXCHANGE RESULTS

The incoming longitudinal emittance was determined by tuning the 9-cell cavity rf phase for minimum energy spread, as observed at the XS3 screen in the straight line spectrometer. At this phase setting, the bunch length was determined at the X09 OTR screen using a Hamamatsu C5680 streak camera operating with synchroscan vertical plugin unit that was phase locked to 81.25 MHz [20]. The standard input optics barrel using transmissive fused silica lenses was replaced with a reflective mirror optics configuration to reduce the contributions of chromatic temporal dispersion to the observed bunch lengths. Typical energy spread and bunch length 1-sigma values are 8-10 keV and 3 ps or 0.9 mm, respectively. Under the assumption of an upright longitudinal phase space, the product of these rms values would give longitudinal emittance.

For the outgoing transverse emittances, the beam size was first determined at X23, and then with the slit assemblies inserted at X23, the slit images were measured at X24. The X23-X24 image pairs were used with a drift of 0.56 m. The outgoing emittances and C-S parameters were calculated with the same MATLAB program used for the input emittances.

The outgoing longitudinal emittance was evaluated by an energy-spread measurement at the XS4 screen in the vertical spectrometer and by the bunch-length measurement at X24 with the OTR transported to the same streak camera as used at X09 and/or with the X24 FIR CTR transported to a Martin-Puplett interferometer [21]. After the exchange, bunch lengths are typically sub-ps and the energy spread is about 6-8 keV. Due to optics constraints and the limitations of the longitudinal measurement (i.e., no deflecting cavity), it was difficult to establish without ambiguity the required upright phase-ellipse conditions so we only have an upper limit formed by the product of the projected components. Examples of the OTR streak image and bunch length profile obtained at X24 with the 5-cell rf cavity off (left) and on (right) are shown in Fig. 7. The initial bunch length sigma was 2.1 ps and the EEX one was ~0.7 ps. Fig. 8 shows the tracking of 25 shots with the 5-cell cavity off (red dots) and 5-cell cavity on (blue squares). The results are quite consistent, but the scatter of points is larger in the cavity-on data since the corrections to the raw data are fractionally larger than for the cavity-off data. The intrinsic camera resolution is 0.6 ps for a red laser source.

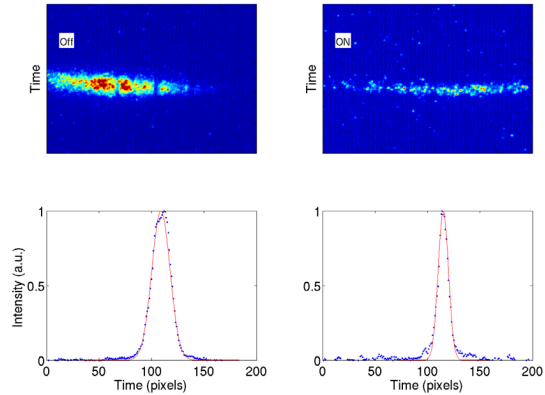

Figure 7: Initial comparison of the OTR streak images taken at X24 with the 5-cell cavity power off (left) and on for EEX (right). The projected temporal profiles (blue dots) are shown below the images, respectively. The red curve is the Gaussian fit to the data profile in each.

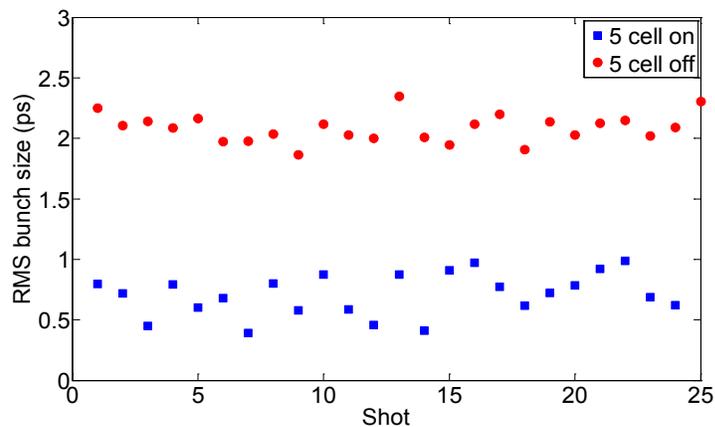

Figure 8: Comparison of the streak camera results at X24 for the 5-cell cavity off (red dots) and on (blue squares) when the exchange was tuned for minimum delta t.

We proceeded to test emittance exchange, and took advantage of the newly installed YAG:Ce crystal at X24 as well to measure the outgoing transverse parameters. The initial values are at the left and the

outgoing values at the right for the EEX results presented below:

$$\varepsilon^n{}_x = 2.6 \pm 0.3 \text{ mm mrad} \rightarrow 11 \pm 2 \text{ mm mrad}$$
$$\varepsilon^n{}_z = 11.5 \pm 1.5 \text{ mm mrad} \rightarrow 5-6 \text{ mm mrad}$$
$$\varepsilon^n{}_y = 2.5 \pm 0.3 \text{ mm mrad} \rightarrow 3.7 \pm 0.3 \text{ mm mrad}$$

The longitudinal-to-transverse(x) exchange (blue arrow) and the transverse(x)-to-longitudinal exchange (purple arrow) are clear, albeit with some growth in the latter. The outgoing longitudinal emittance growth may be explained by a transport matrix model including space charge effects [22], by a 5-cell cavity rf gradient setting error, and by a depth-of-focus effect across the elongated image in the spectrometer. Corrections of the latter two effects in a subsequent test accounted for most of the "growth" as will be reported elsewhere [19]. The vertical emittance growth was about 50%, perhaps due to some coupling between the x-y planes. The statistical errors are calculated in the transverse emittance code [10].

## VI. SUMMARY

In summary, we have investigated the contributions of several terms to beam profile station system resolution at A0PI and converted most of the YAG:Ce powder screens to YAG:Ce single crystal screens. We have also changed the geometry of the screens at the divergence stations and, as of recently, the spectrometer stations to eliminate the compromising depth-of-focus issues across the field of view. The bunch length diagnostics had been previously upgraded [20,21]. These diagnostic upgrades have resulted in more reliable characterization of the transverse and longitudinal emittances before and after EEX (with an almost factor of two reduction in several values compared to the preliminary report of ref. [5] in which no projected size corrections were made and no upgrades had been done) and subsequently led to the long-sought observations of nearly one-to-one exchanges in more recent tests [19].

## ACKNOWLEDGEMENTS

The authors acknowledge the beamline vacuum work of W. Muranyi and B. Tennis and support from M. Wendt, M. Church, E. Harms of Fermilab. They also acknowledge K. Floettmann of DESY for providing the YAG:Ce powder screens and P. Piot of NIU and F. Stephan of PITZ for discussions on their application.
## REFERENCES

[1]  M. Cornacchia and P. Emma, Phys. Rev. ST Accel. Beams **5**, 084001 (2002).

[2]  P. Emma, Z. Huang, and K.-J. Kim, Phys. Rev. ST Accel. Beams **9**, 100702 (2006).

[3]  J. P. Carneiro *et al.*, Phys. Rev. ST Accel. Beams **8**,040101 (2005).   *Proc. of PAC99, 2027 (1999).

[4] T.W. Koeth, "An Observation of a Transverse to Longitudinal Emittance Exchange at the Fermilab A0 Photoinjector", (Ph. D. thesis), Rutgers University, (2009).

[5] T. W. Koeth *et al.*, Proceedings of the 2009 Particle Accelerator Conference, Vancouver BC, (PAC09), TU4PBI01, (2009).

[6] A. H. Lumpkin *et al.*, Proceedings of the 2009 Free Electron Laser Conference, Liverpool UK, (FEL09), TUPC46, (2009).

[7]  B. X. Yang (private communication).

[8] M. Maesaka *et al.*, Proceedings of the 10[th] European Workshop on Beam Diagnostics and Instrumentation for Particle Accelerators, Basel Switzerland, (DIPAC09), MOOA03,(2009).

[9] C.H. Wang et al., International Conference on Accelerator and Large Experimental Physics Control Systems, Trieste Italy, (ICALEPS), 284, (1999).

[10] R. Thurman-Keup et al., "Transverse Emittance and Phase Space Program Developed for Use at the Fermilab A0 Photoinjector", Proceedings of the 2011 Particle Accelerator Conference, New York, NY, (PAC11), to be published, (2011).